\documentclass[slac_one]{revtex4}
\usepackage{graphicx}
\usepackage{fancyhdr}
\begin{document}

\title{{\small{Hadron Collider Physics Symposium (HCP2008),
Galena, Illinois, USA}}\\ 
\vspace{12pt}
Geant4 validation with CMS calorimeters test-beam data} 

%

\author{Stefan Piperov}
\affiliation{Inst. Nuclear Res. and Nuclear Energy, Bulgarian Acad. Sci., BG-1784 Sofia, Bulgaria}
\affiliation{FNAL, Batavia, IL 60510, USA}

\begin{abstract}
CMS experiment is using Geant4 for Monte-Carlo simulation of the detector
setup. Validation of physics processes describing hadronic showers is a major concern
in view of getting a proper description of jets and missing energy for
signal and background events. This is done by carrying out an
extensive studies with test beam using the prototypes or real detector
modules of the CMS calorimeter. These data are matched with Geant4
predictions. Tuning of the Geant4 models is carried out and steps to be 
used in reproducing detector signals are defined in view of measurements 
of energy response, energy resolution, transverse and longitudinal shower
profiles for a variety of hadron beams over a broad energy spectrum
between 2 to 300 GeV/c.
\end{abstract}

\maketitle

\thispagestyle{fancy}


\section{INTRODUCTION} 
Series of Test-Beam measurements have been performed on the calorimetric system 
of CMS over the last few years in order to optimize its design and study its 
performance. Detailed Monte-Carlo simulations of the test-beam configuration  
have been made and the results compared with these test-beam measurements, 
leading to a number of improvements in the simulation code. 
Presented here is a comparison between the results from the 2006 Test-Beam and the 
Geant4-based Monte-Carlo simulations of this setup.

\section{THE COMPACT MUON SOLENOID AND ITS CALORIMETRIC SYSTEM}
\begin{figure}
 \begin{tabular}{cc}
  \includegraphics[width=0.45\linewidth]{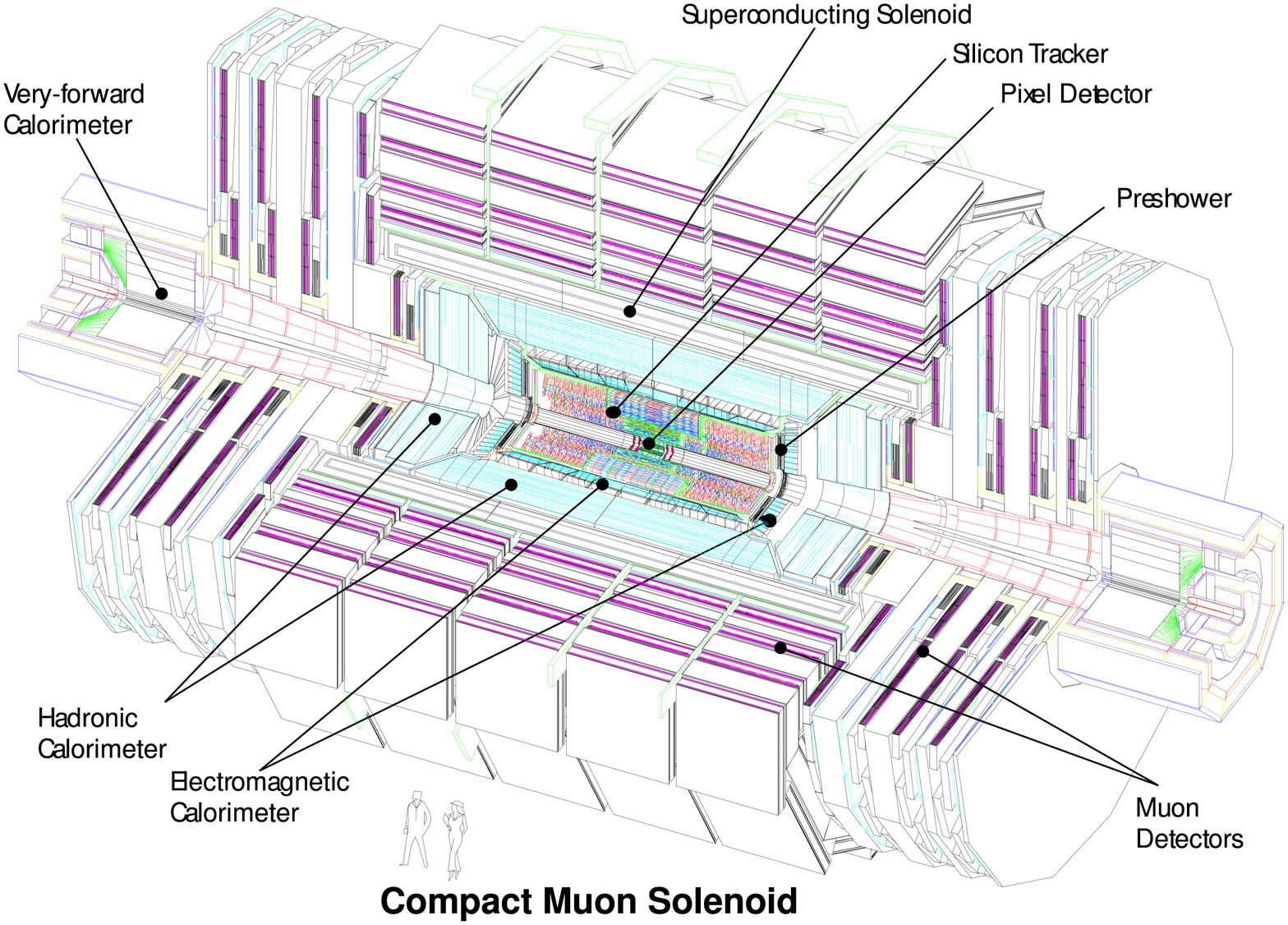}
&
    \includegraphics[width=0.45\linewidth]{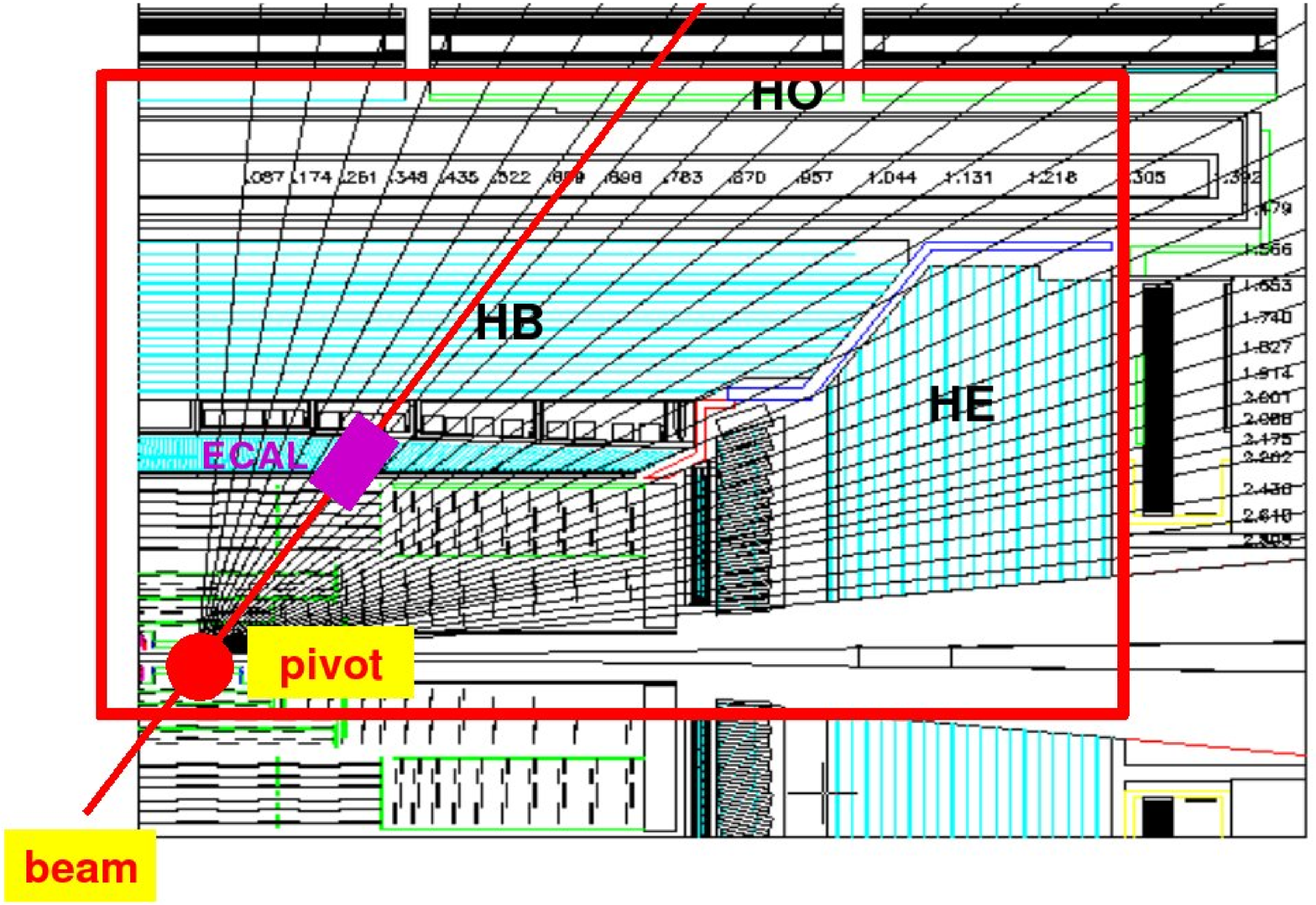}
 \\
a) & b) \\
 \end{tabular}
 \caption{a) CMS experiment with the main subdetectors; 
          b) Calorimetric systems present on main moving table of Test-beam 2006 (framed). 
             Pivot point corresponds to beam-crossing point in CMS; }
   \label{CMS_calorimeters}
\end{figure}

\begin{figure}
 \begin{tabular}{cc}
    \includegraphics[width=0.4\linewidth]{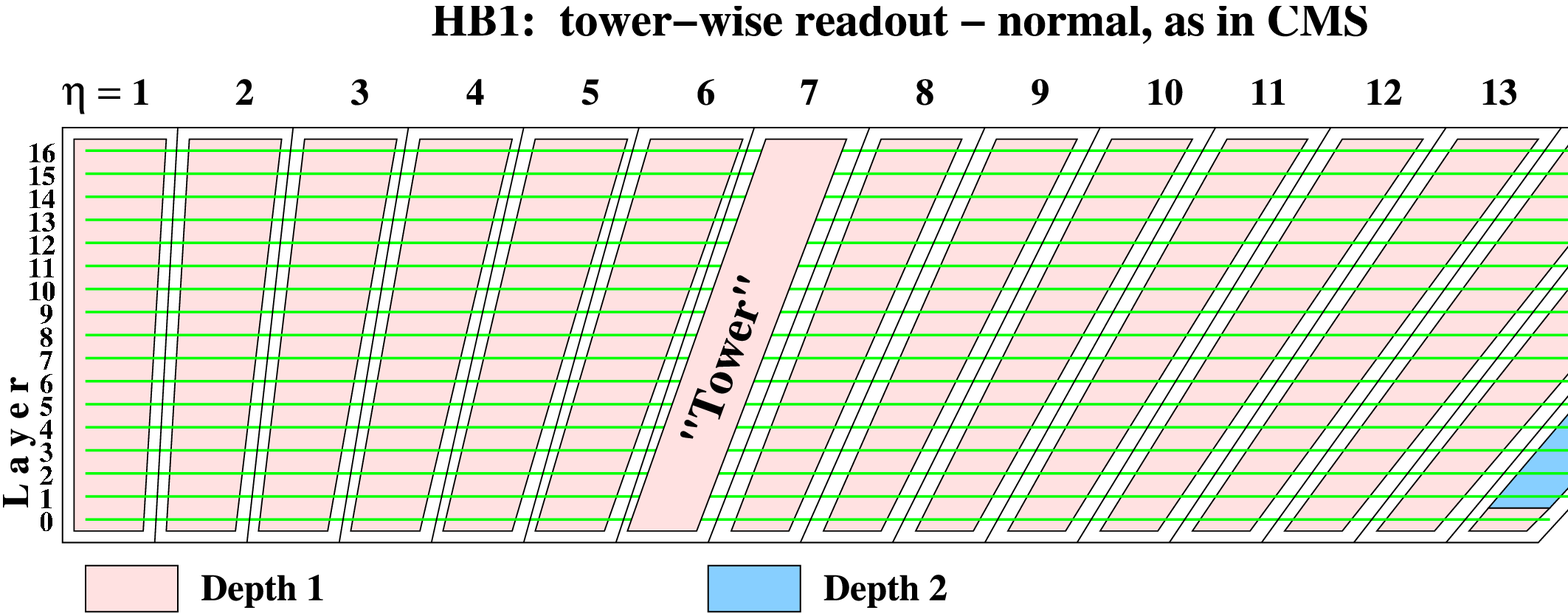}
&
    \includegraphics[width=0.4\linewidth]{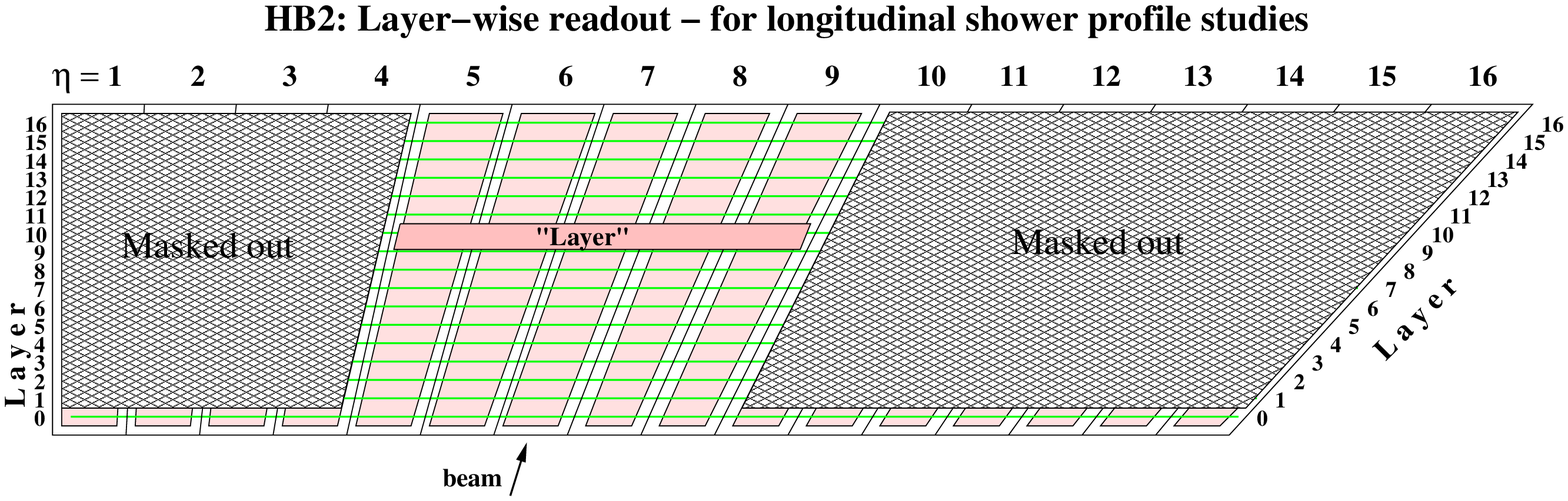}
\\
a) & b) \\
  \end{tabular}
 \caption{Two different read-out schemes for the HB wedges: Tower-wise (a) and Layer-wise (b)}
   \label{HB_readout}
\end{figure}

\begin{figure}
  \includegraphics[width=0.6\linewidth]{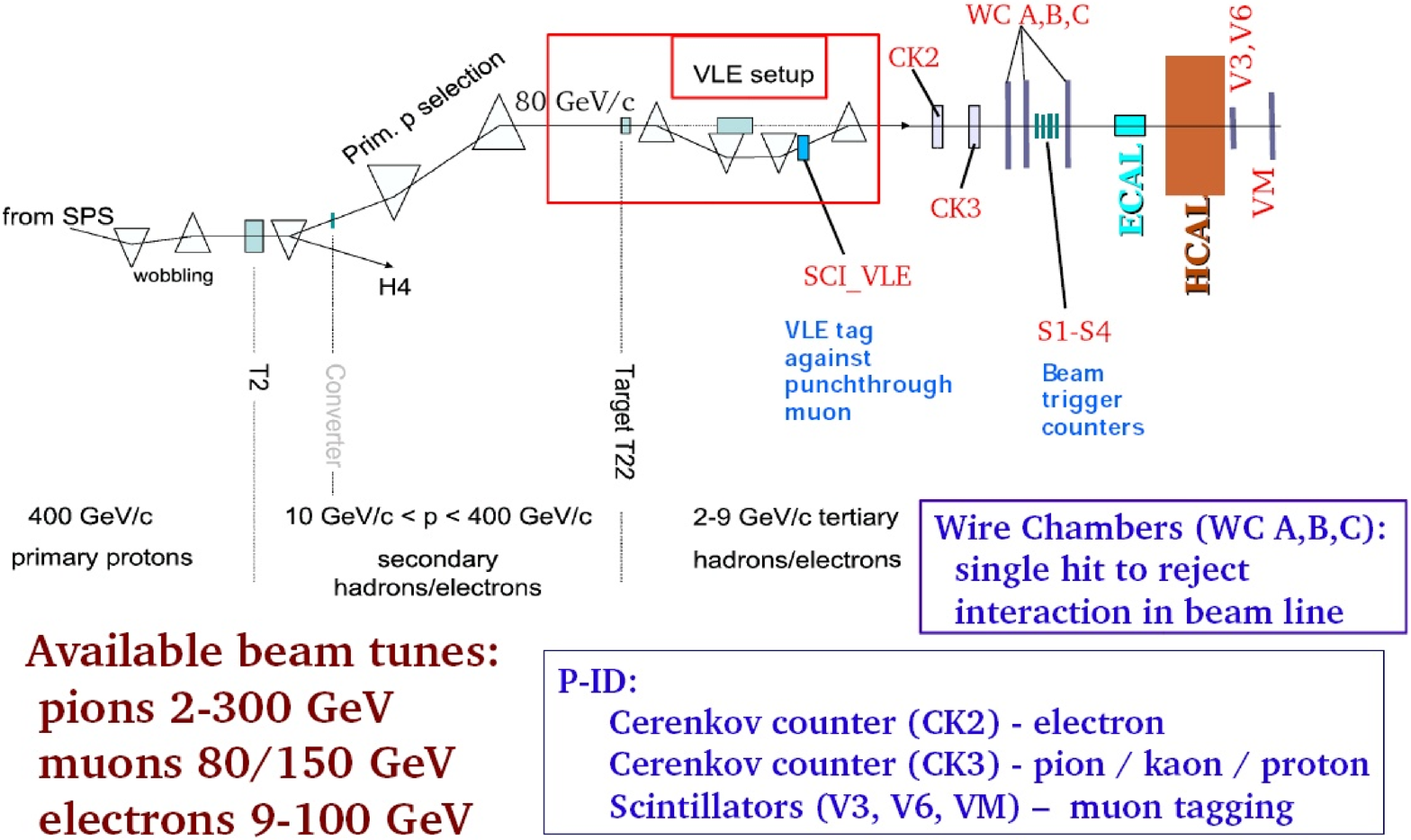}
  \caption{SPS's H2 beam-line. The VLE section is used for production/selection of very low beam energies 
(2 - 10 $GeV$)}
  \label{H2}
\end{figure}

The Compact Muon Solenoid detector (CMS)\cite{CMS:TP} is one of the general purpose detectors
for the Large Hadron Collider (LHC) at CERN, to start operations in 2008.  It's calorimetric 
system consists of Electromagnetic Calorimeter (ECAL) and Hadronic Calorimeter (HCAL) positioned 
inside the superconducting solenoid, and a pair of Very-Forward Calorimeters positioned outside 
of the return yoke of the magnet (fig.\ref{CMS_calorimeters}a). For the purpose of this study
only one super-module of ECAL crystals, without the preshower in front, and a pair of HCAL barrel 
wedges were used, as shown in fig.\ref{CMS_calorimeters}b. The Outer Calorimeter (HO) was present 
in both the test-beam environment and in the simulation geometry, but was intentionally left out
of the comparisons, as its inclusion introduces additional uncertainties which were not well 
understood at the time of this study. The Forward Calorimeter (VF) was not included in the study.
One of the two HB wedges used in the test-beam was wired in the normal "tower-wise" way (Fig.\ref{HB_readout}a), while the 
second was wired in a "layer-wise" manner (Fig.\ref{HB_readout}b) to allow for the studies of longitudinal shower profiles.
Fig.\ref{H2} shows the H2 beam-line arrangement at CERN's SPS accelerator. The VLE section enables the formation of 
very-low-energy beam tunes, allowing beam momenta as low as 2GeV/c to be studied.

\section{THE MONTE-CARLO SIMULATION}
\begin{figure}
  \includegraphics[width=0.8\linewidth]{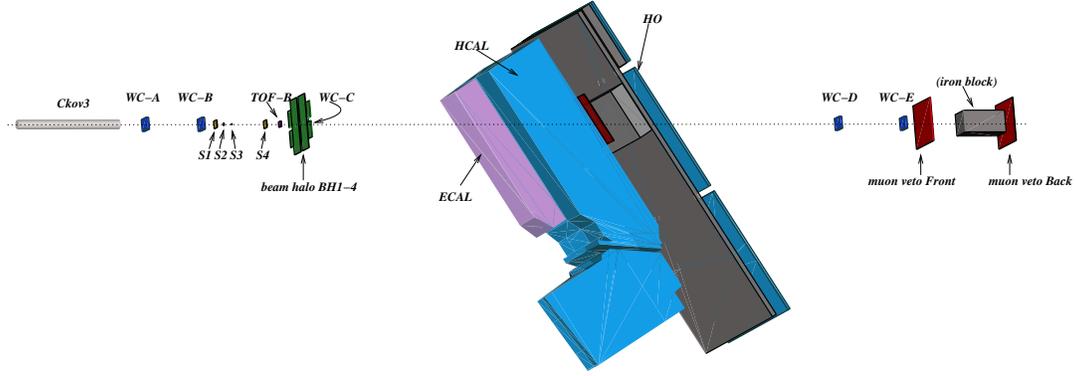}
  \caption{Detailed MC simulation geometry }
  \label{MCH2}
\end{figure}

The full simulation of the test-beam setup was performed using the CMSSW software frameworks, 
which uses internally the Geant4 toolkit\cite{Agostinelli:2002hh},\cite{Allison:2006ve}. Fig.\ref{MCH2} shows an overview of the detailed geometry
used in the simulation. All beam-line elements were carefully described and their responses simulated.

Several releases of the Geant4 code were used in the comparisons, and with each of them several physics-lists describing the 
interaction of hadrons with matter were tested. Historically the LHEP physics list was used as a default, but it's limitations at
high particle momentum warranted it's replacement with the QGSP physics list. Physics lists using the Bertini Cascade became 
available recently, bringing in new potential for better describing the interactions at low energies, and were used in this study as well.
Most results in this paper are from release 9.1.p02 of Geant4.

\section{RESULTS}
\begin{figure}
 \begin{tabular}{cc}
  \includegraphics[width=0.33\linewidth]{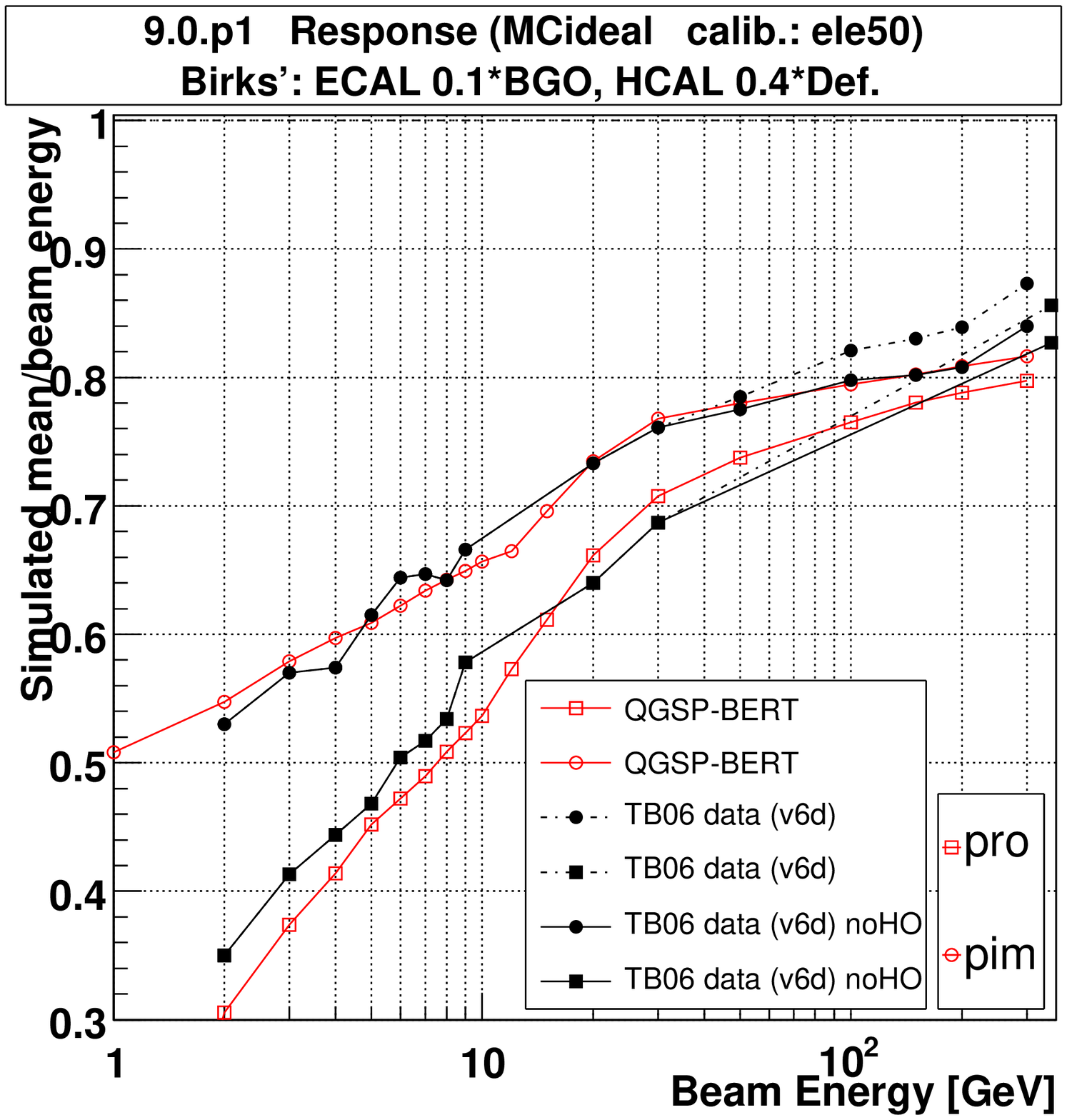}
&
    \includegraphics[width=0.33\linewidth]{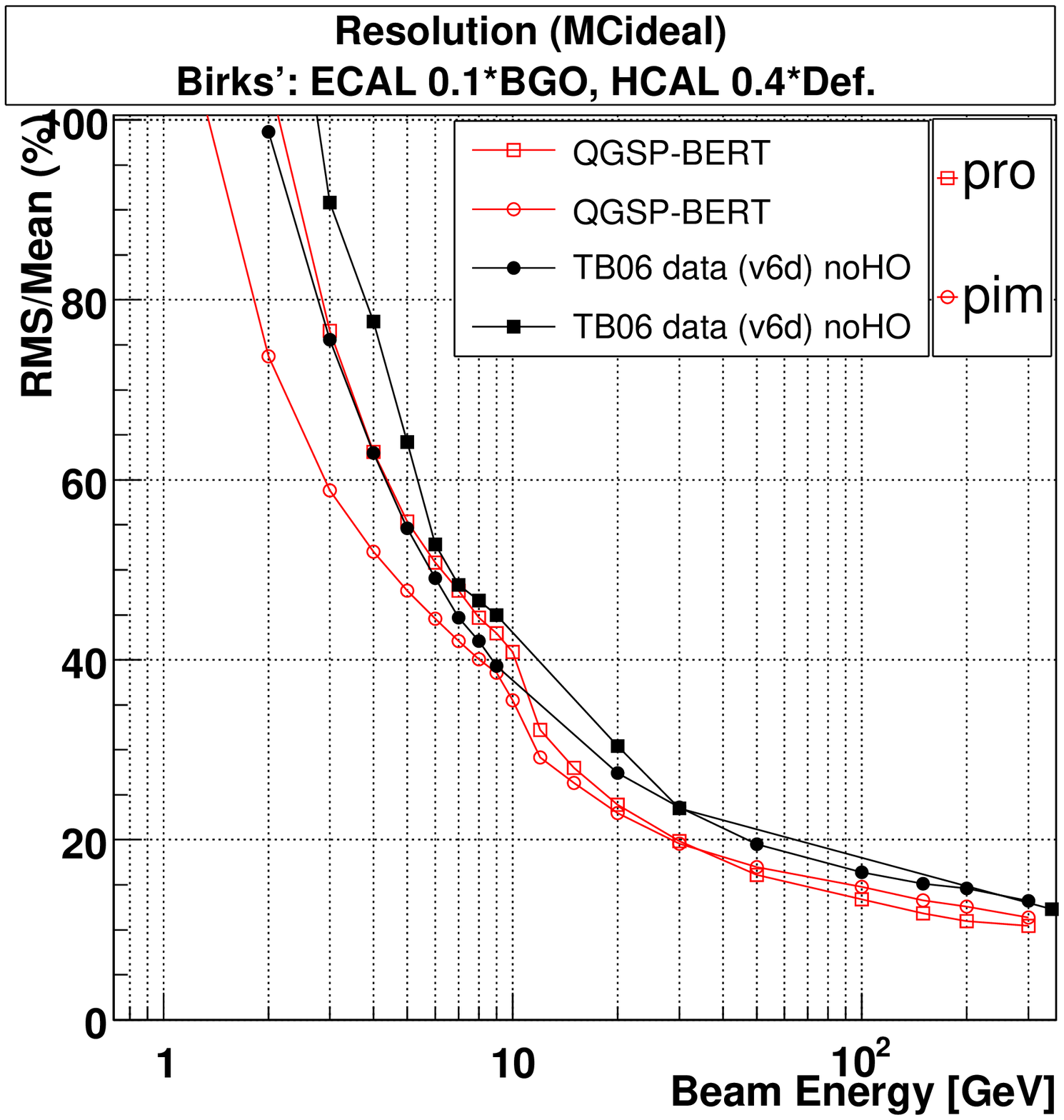}
 \\
  \includegraphics[width=0.33\linewidth]{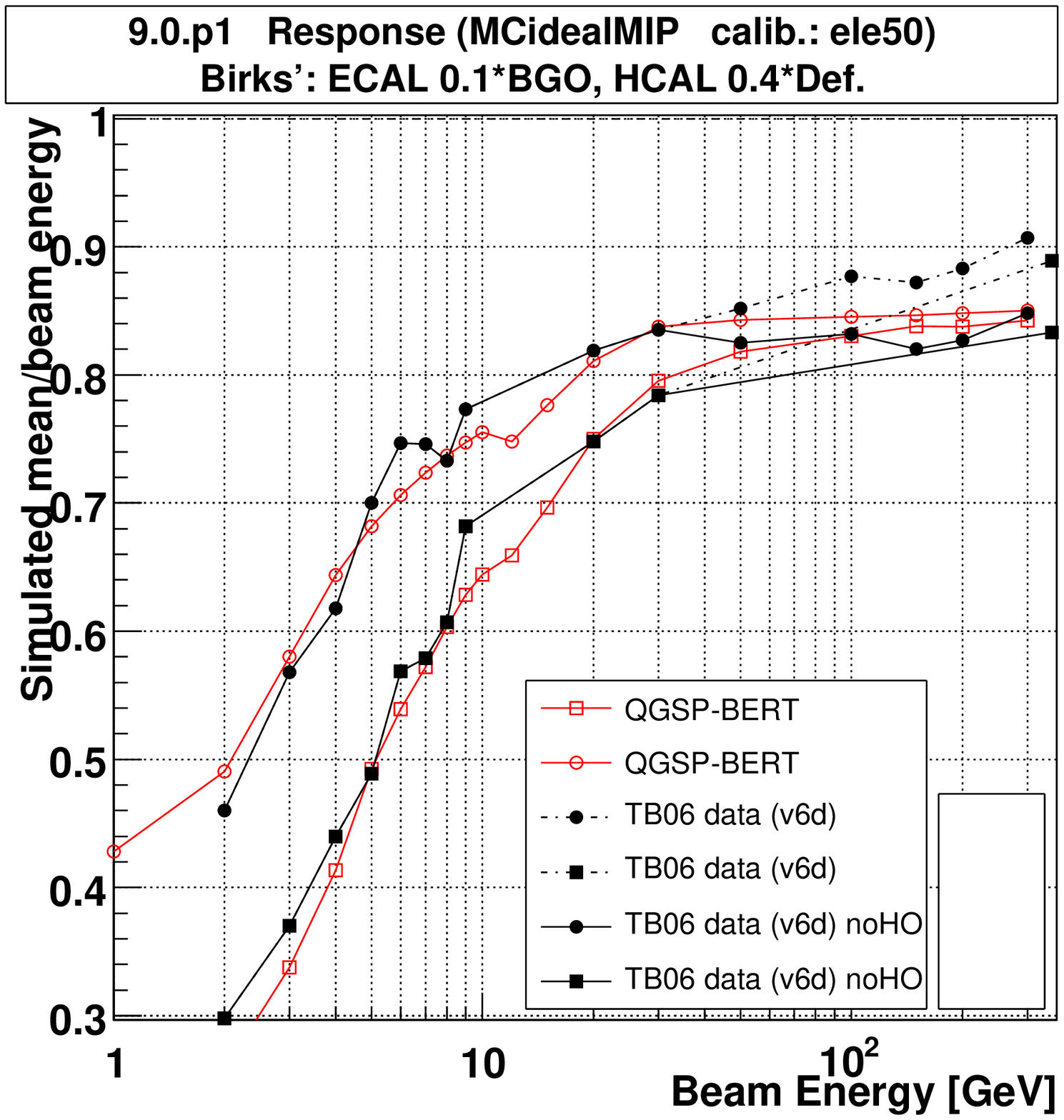}
&
    \includegraphics[width=0.33\linewidth]{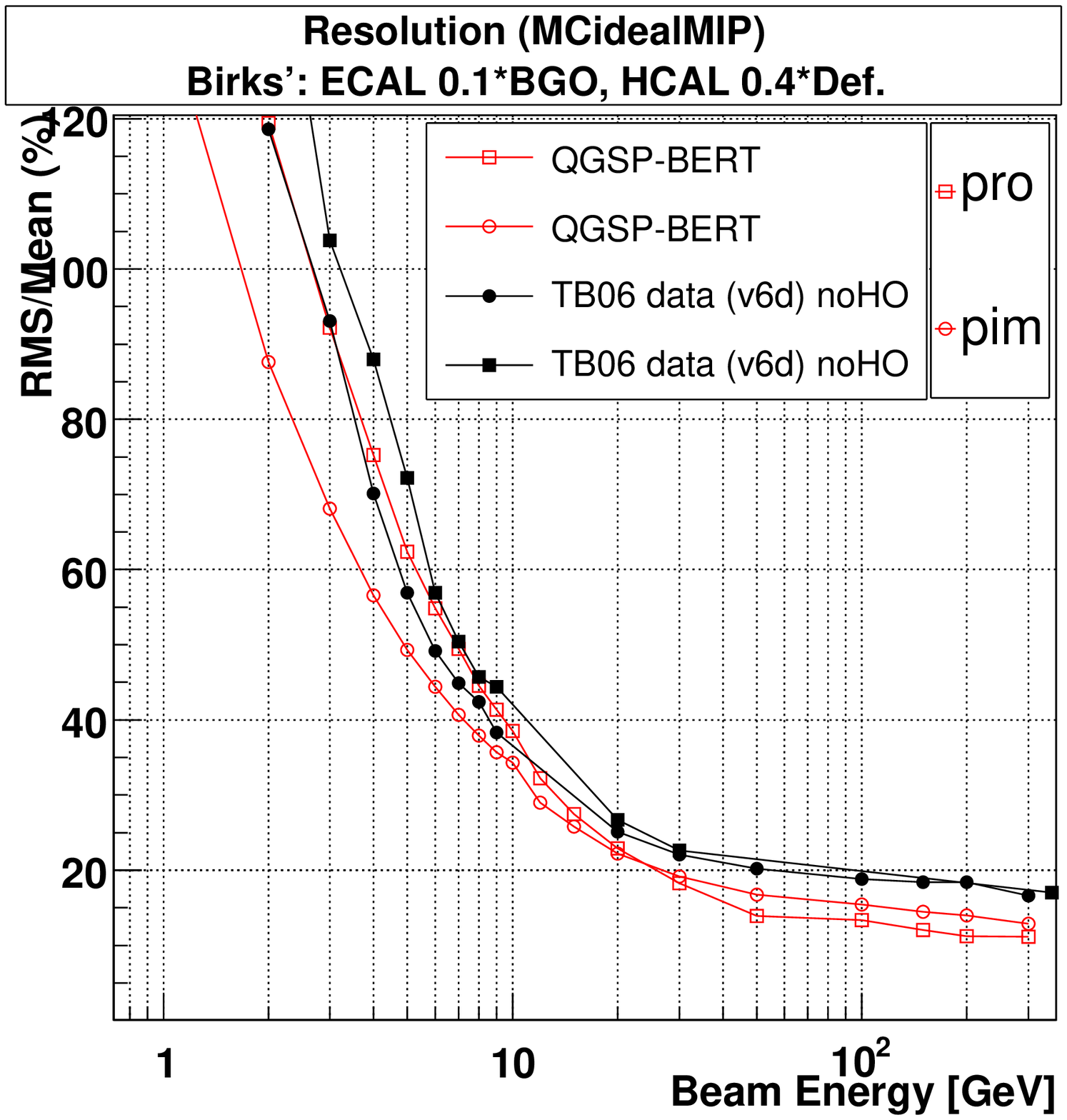}
 \\
 \end{tabular}
 \caption{ Response (left) and resolution (right) of the simulated calorimertic system compared to the measured in TB2006 for the case of combined ECAL+HCAL (top) and HCAL alone (bottom).}
   \label{response_and_resolution}
\end{figure}

\begin{figure}
 \begin{tabular}{ccc||ccc}
  \includegraphics[width=0.2\linewidth]{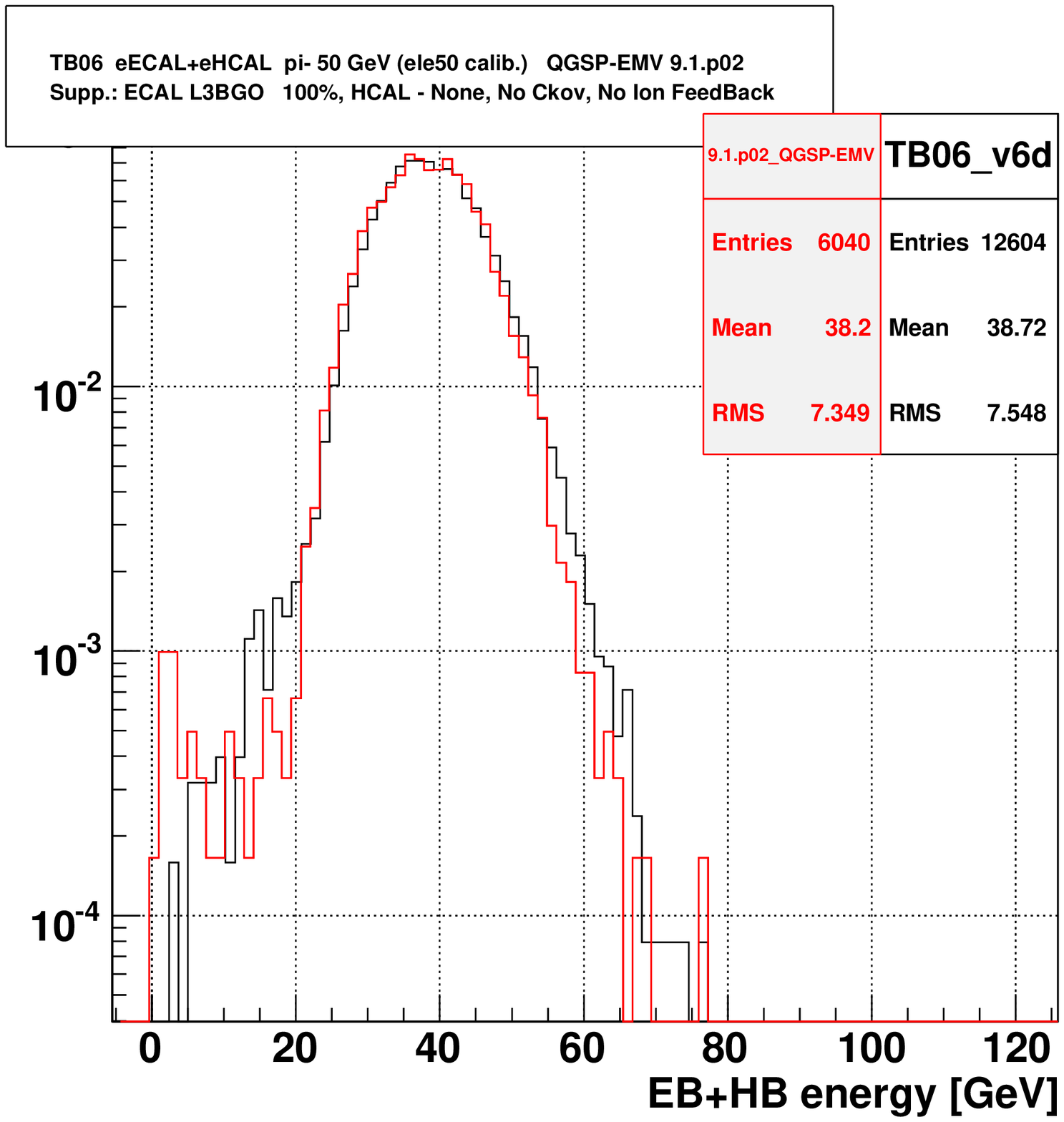}
&
  \includegraphics[width=0.2\linewidth]{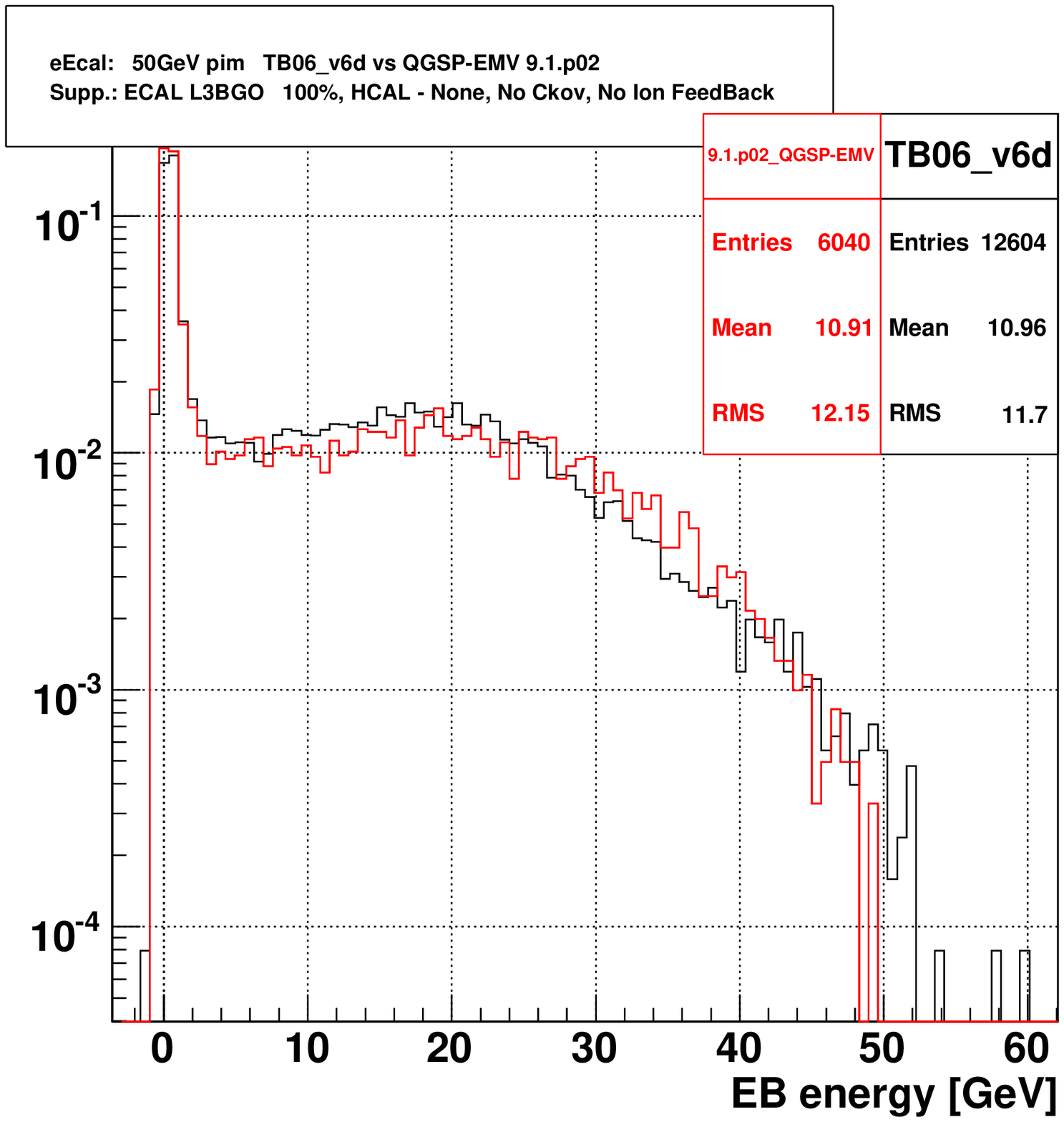}
&
  \includegraphics[width=0.2\linewidth]{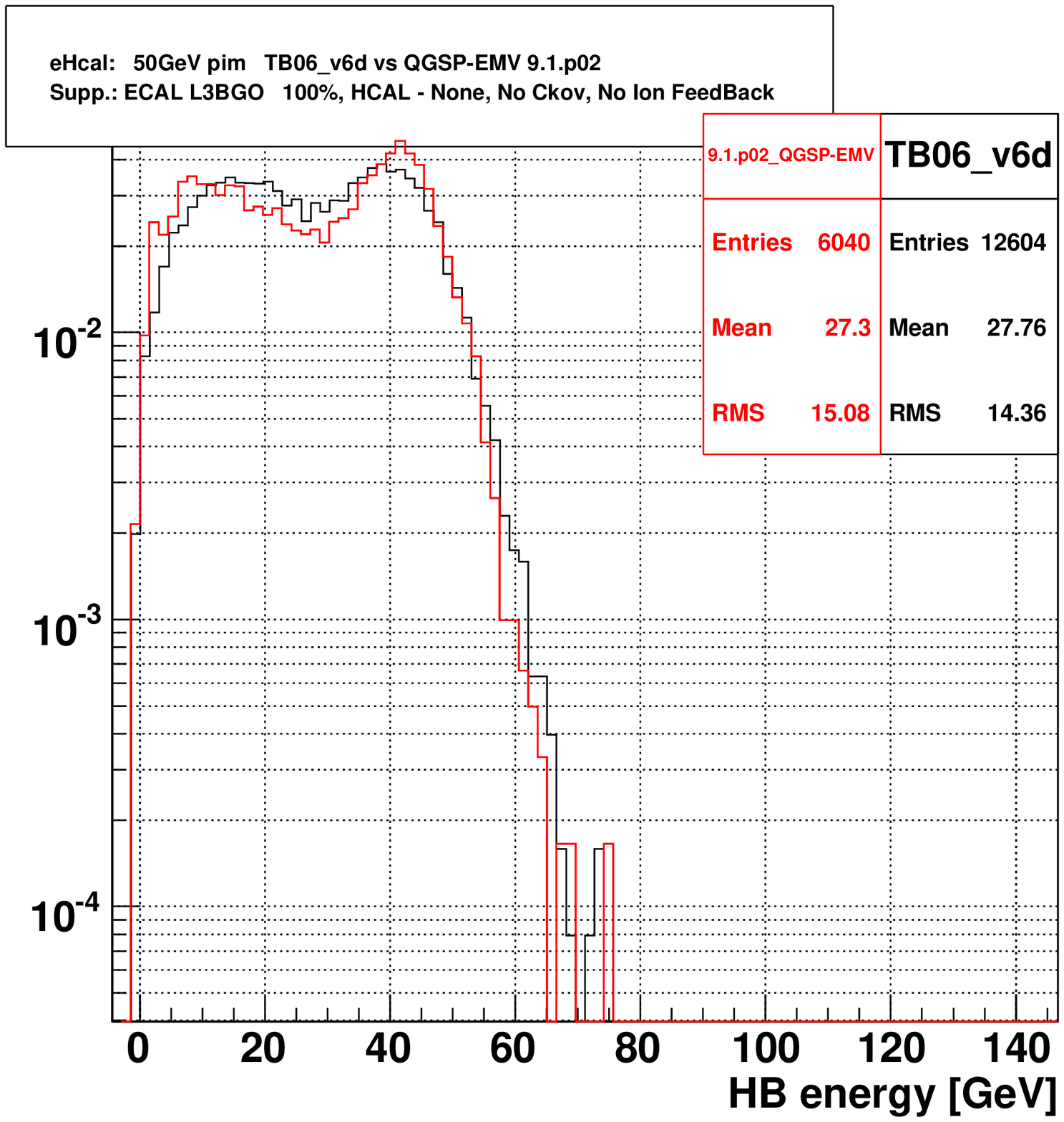}
& &
  \includegraphics[width=0.2\linewidth]{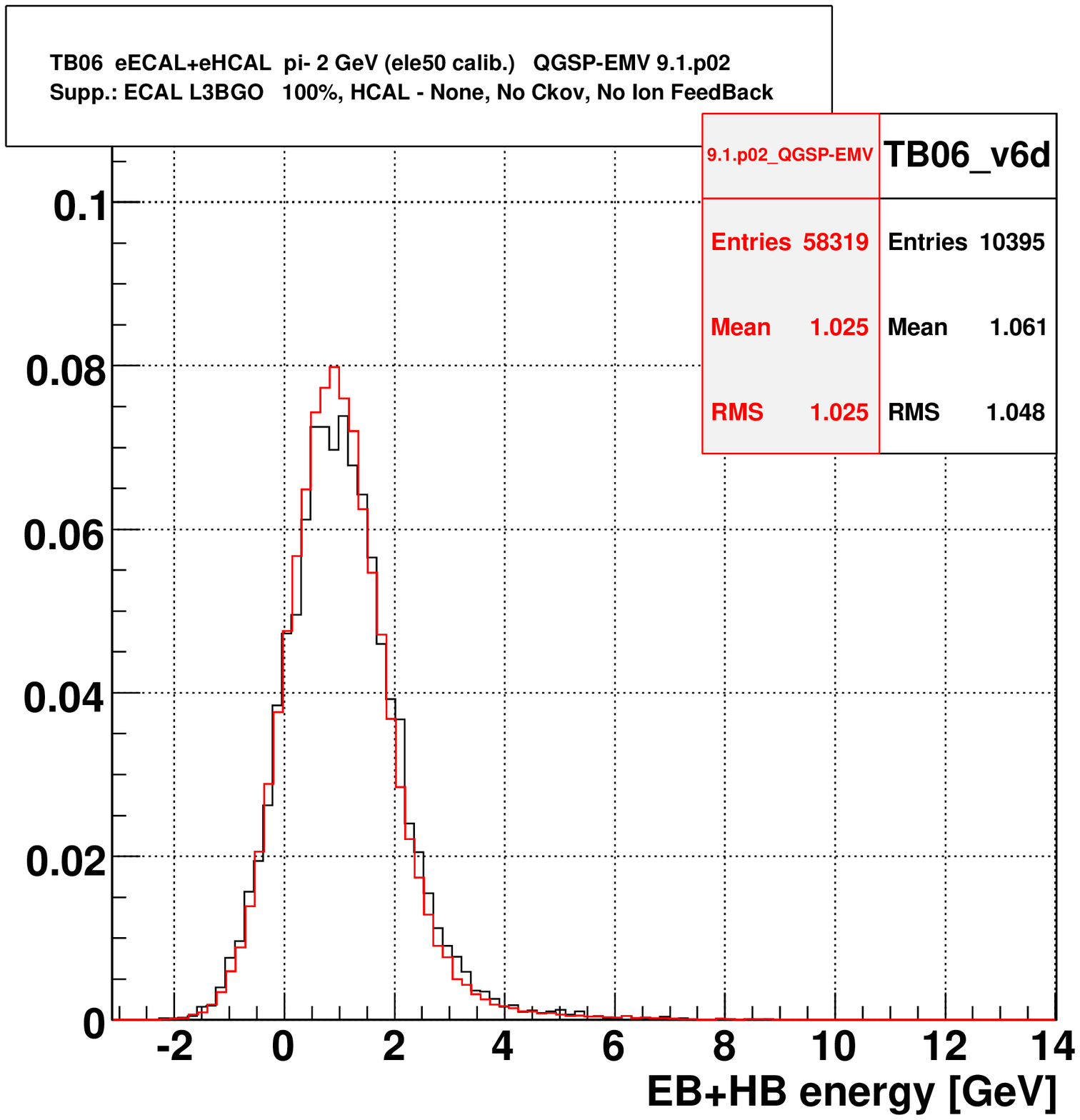}
&
  \includegraphics[width=0.2\linewidth]{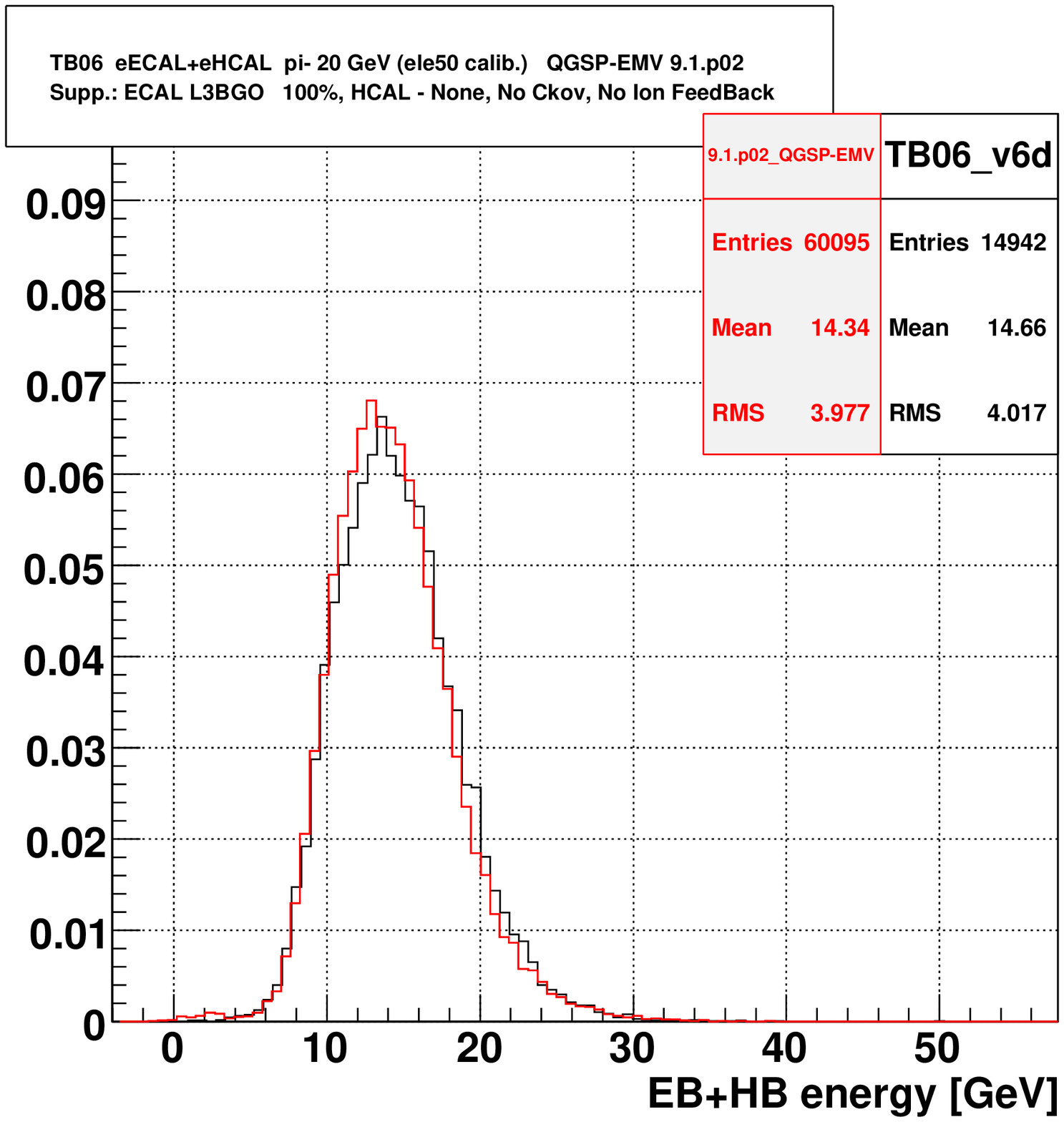}
 \\
 & & & & a) & b) \\
  \includegraphics[width=0.2\linewidth]{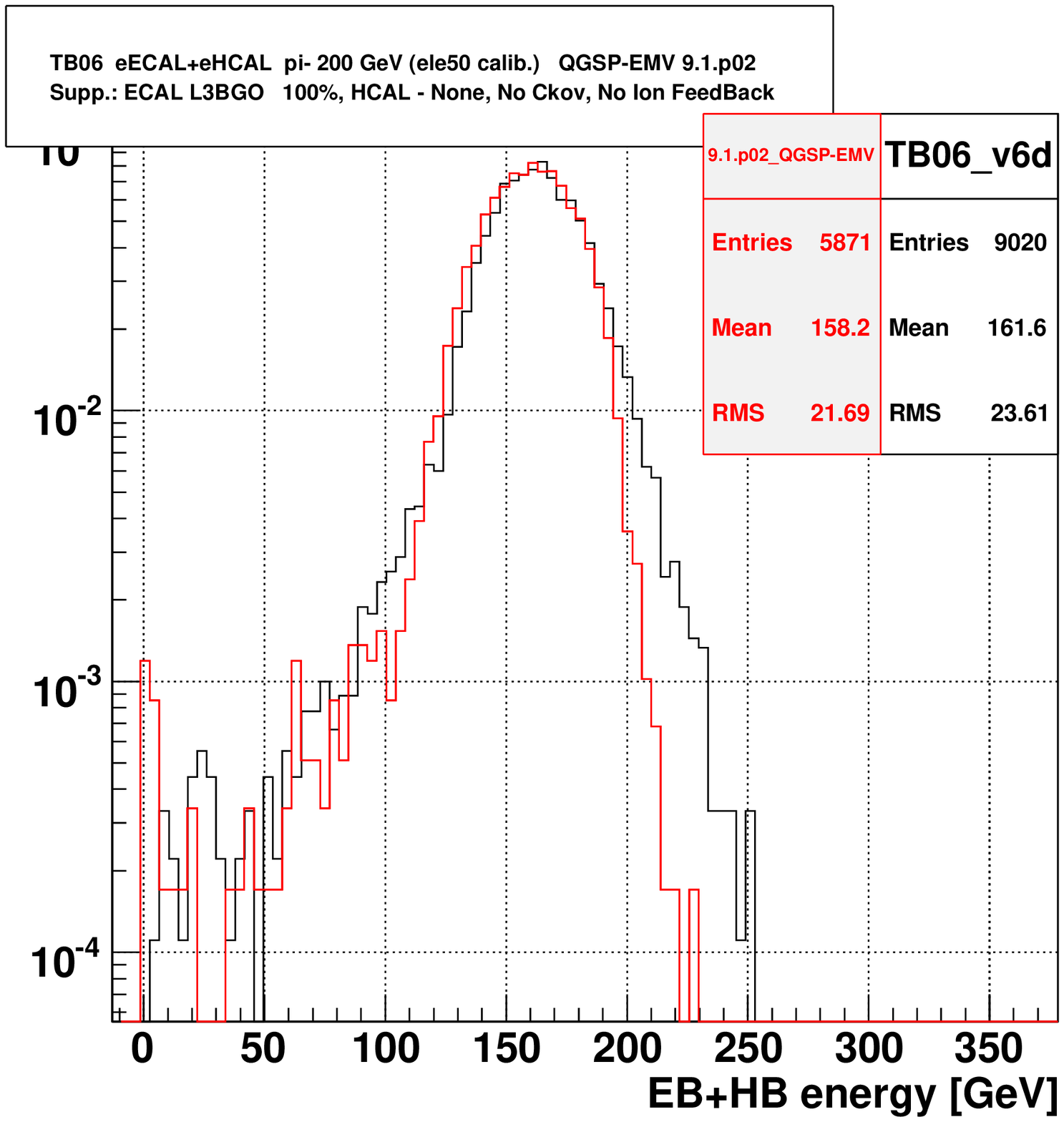}
&
  \includegraphics[width=0.2\linewidth]{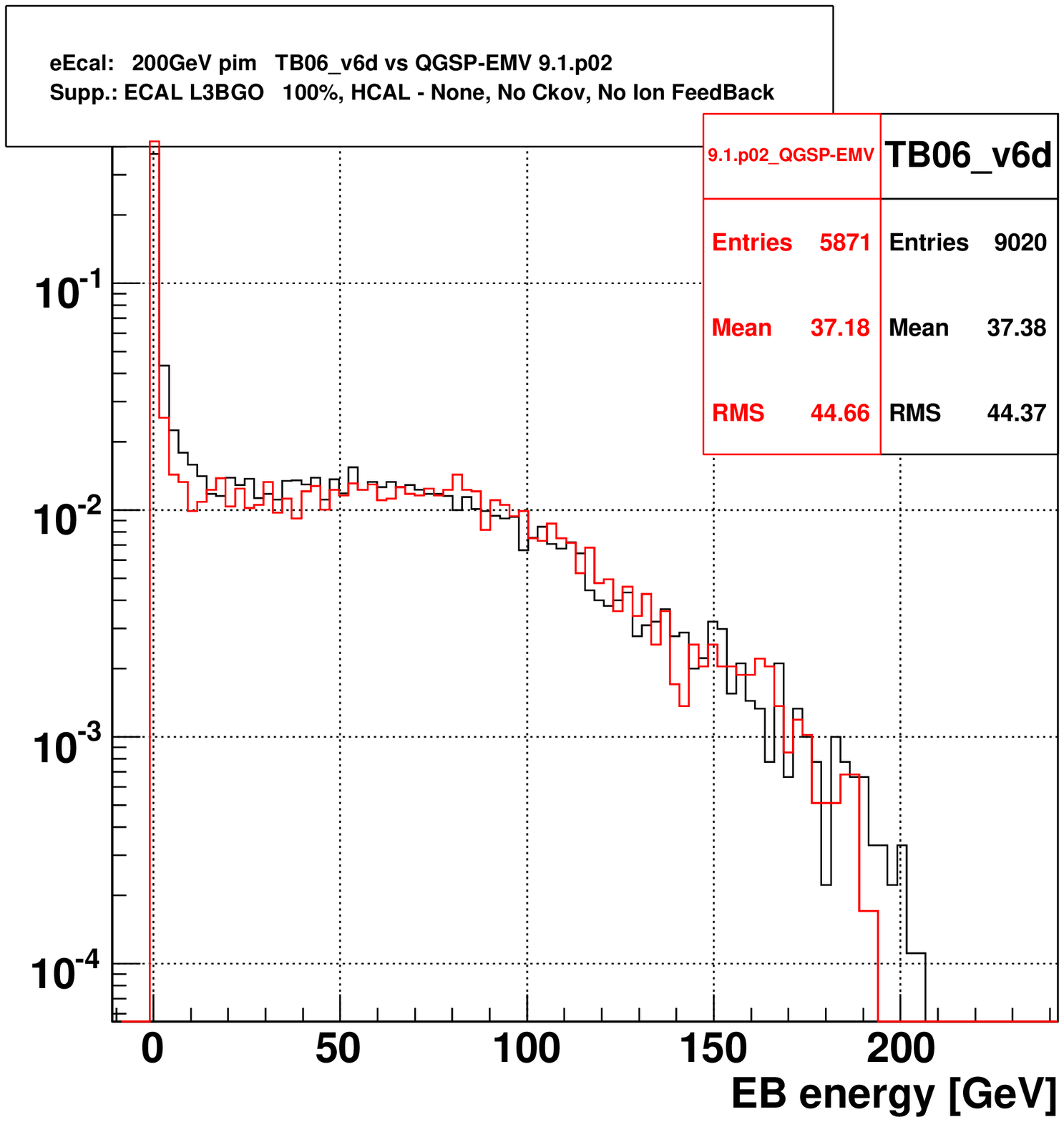}
&
  \includegraphics[width=0.2\linewidth]{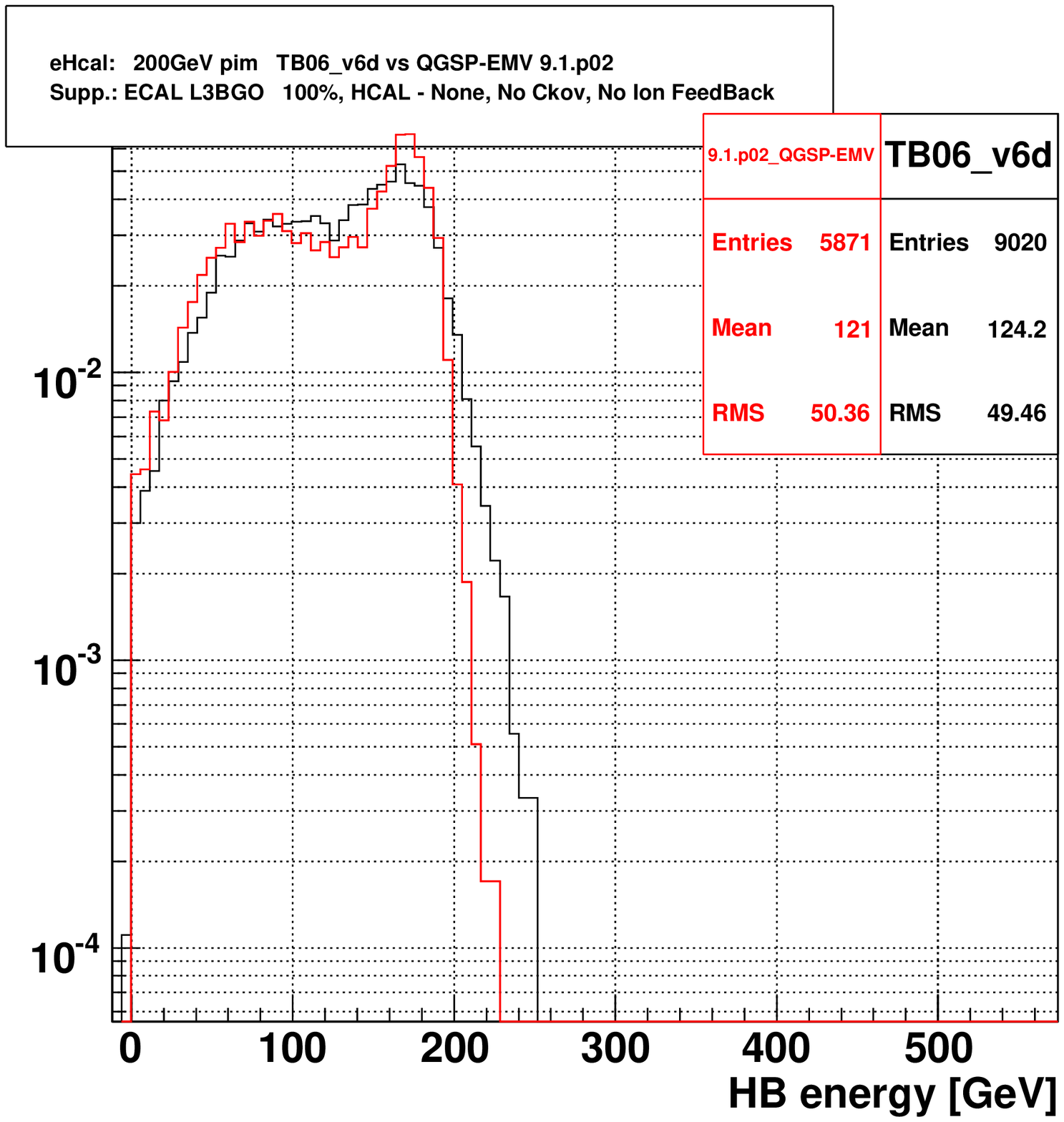}
& &
  \includegraphics[width=0.2\linewidth]{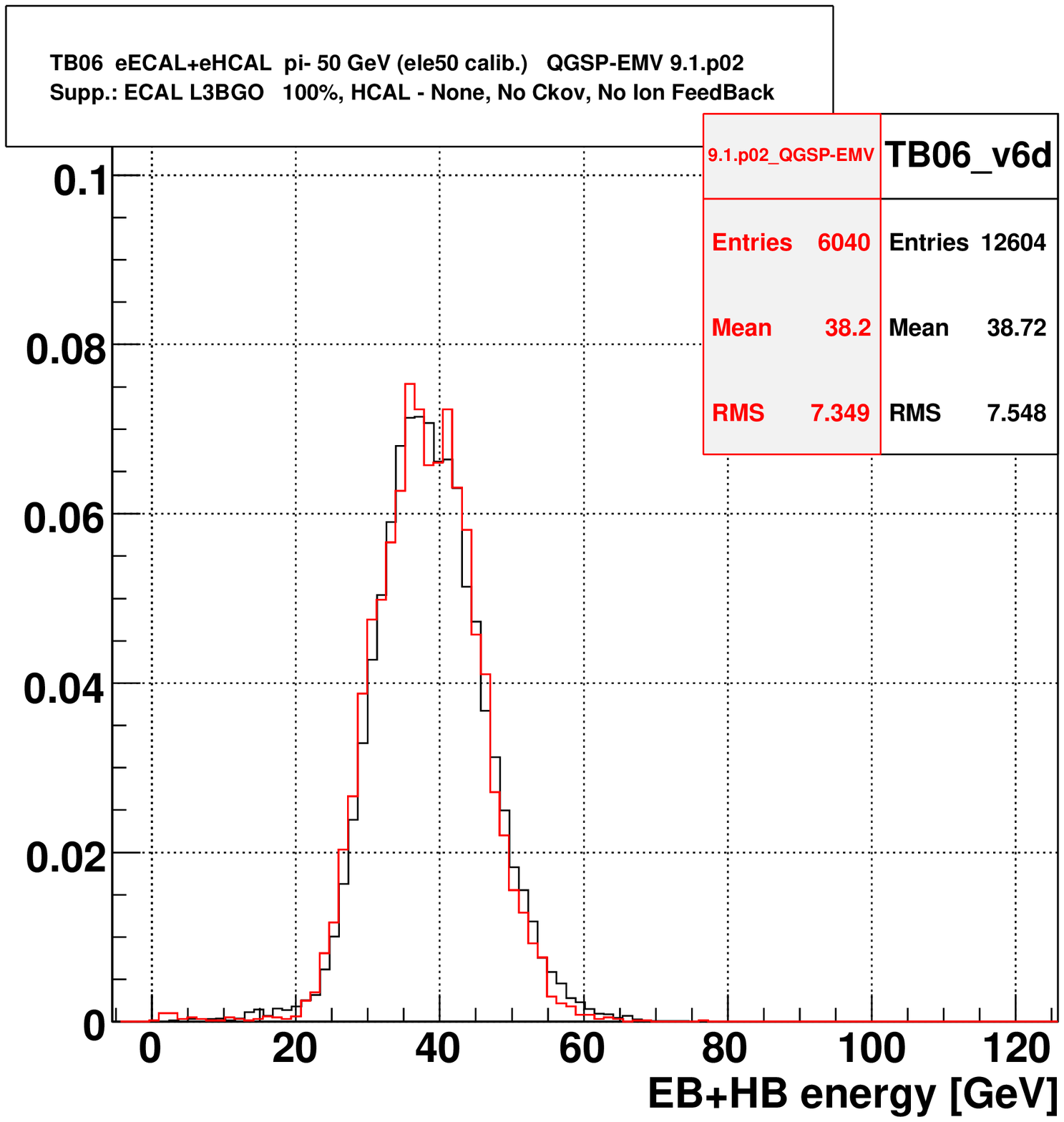}
&
  \includegraphics[width=0.2\linewidth]{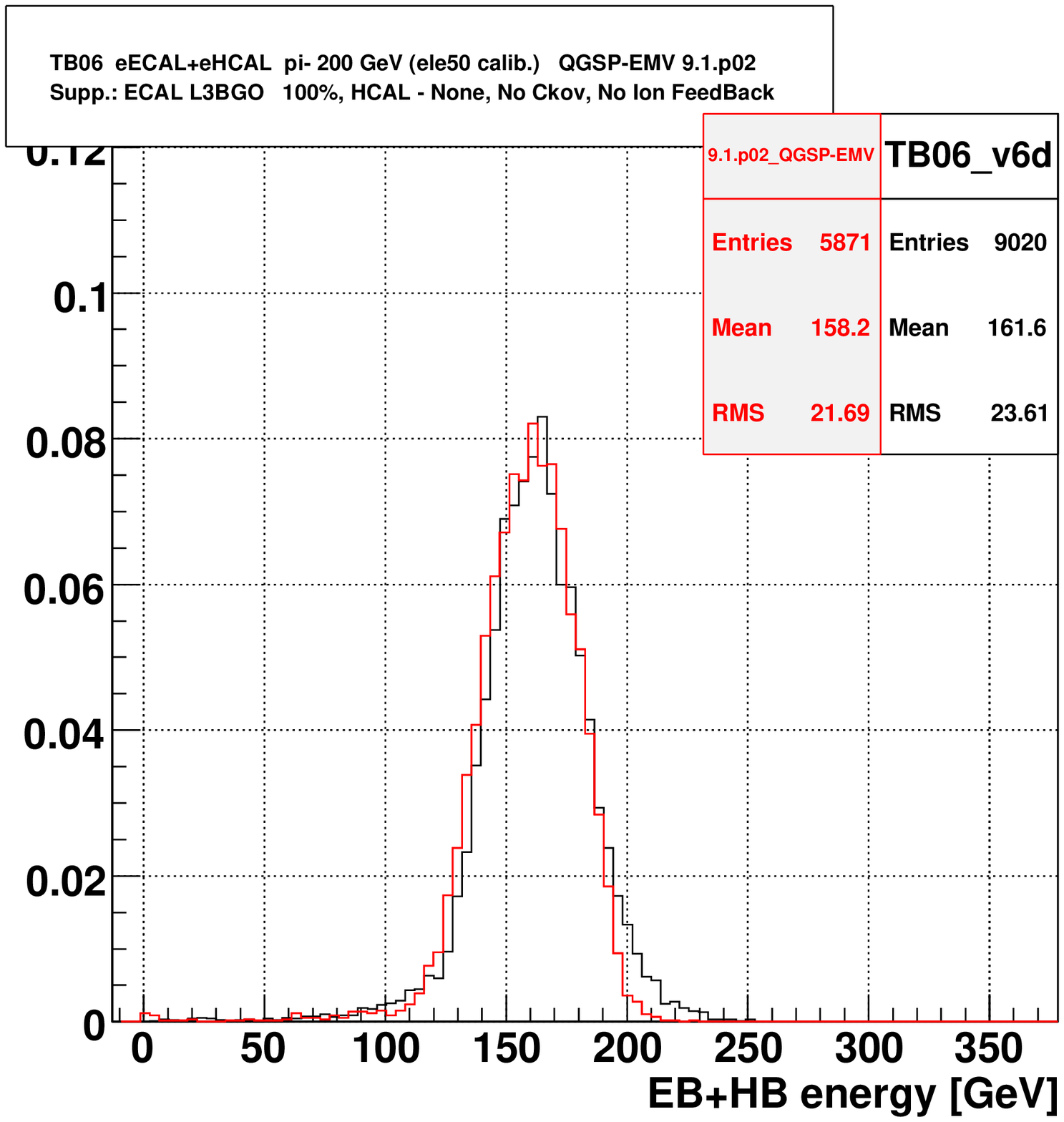}
 \\
 & & & & c) & d) \\
\multicolumn{3}{p{0.6\linewidth}}{ ECAL+HCAL (left);  ECAL (center); HCAL (right);  50GeV (top); and 200GeV (bottom)} & &\multicolumn{2}{p{0.4\linewidth}}{ECAL+HCAL system for 2GeV (a); 20GeV (b); 50GeV (c); and 200GeV (d); pion beams}\\
 \end{tabular}
 \caption{Reconstructed energy spectra of simulated events compared to TB2006 data}
   \label{Log_spectra}
\end{figure} 
\subsection{Linearity of response and Resolution}
Fig.\ref{response_and_resolution} shows the comparison of linearity of response and resolution for the combined system ECAL+HCAL, as well as for the HCAL alone. Overall the agreement is good, except for resolution of the simulated detector being significantly better at high energies.
The source of that discrepancy can be identified in the significantly narrower distribution of HCAL signal at high energies in Fig.\ref{Log_spectra}.
However, the general agreement of the simulated spectra and the measured ones is clear.
\begin{figure}
 \begin{tabular}{ccc}
  \includegraphics[width=0.21\linewidth]{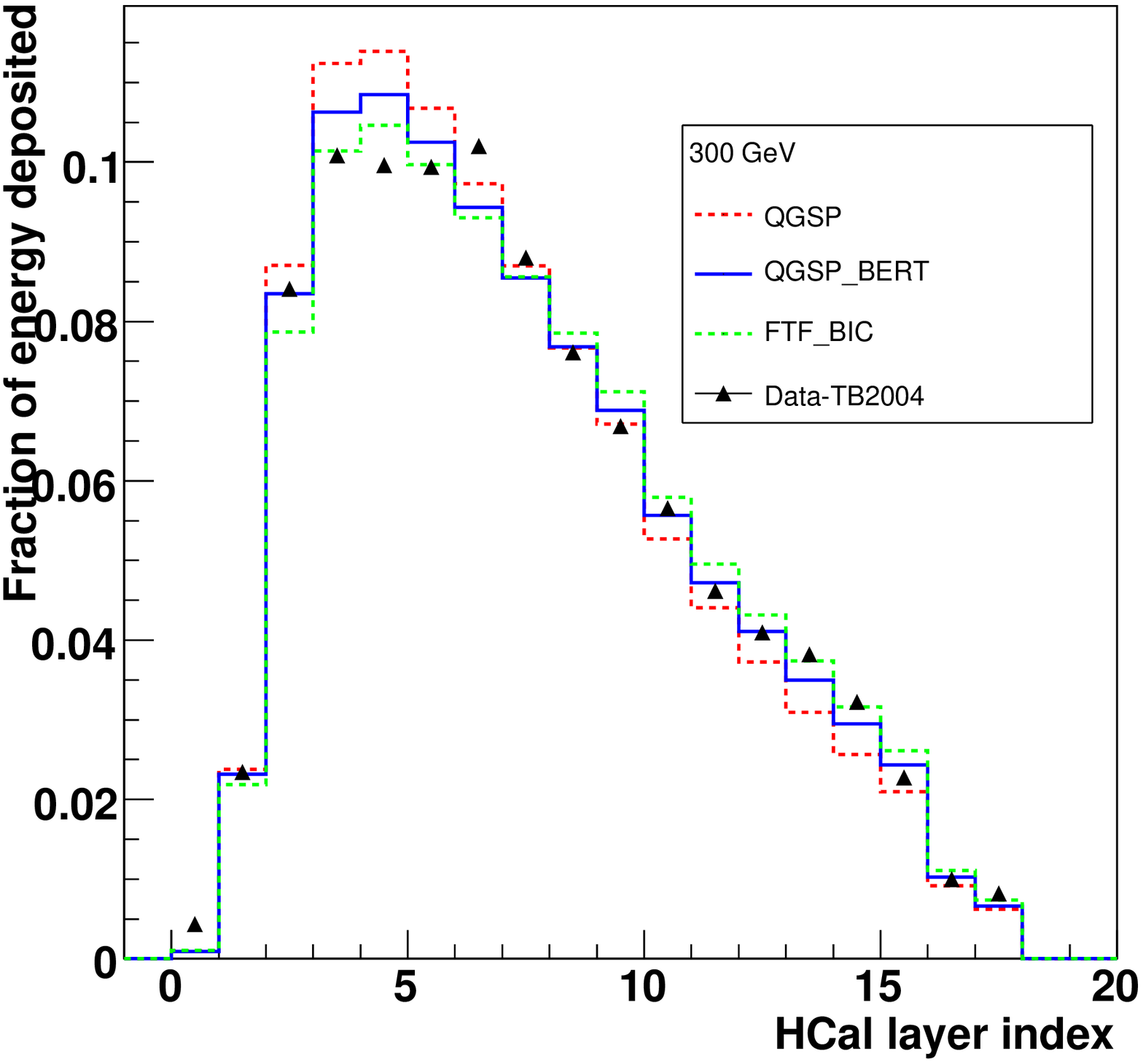}
&
  \includegraphics[width=0.21\linewidth]{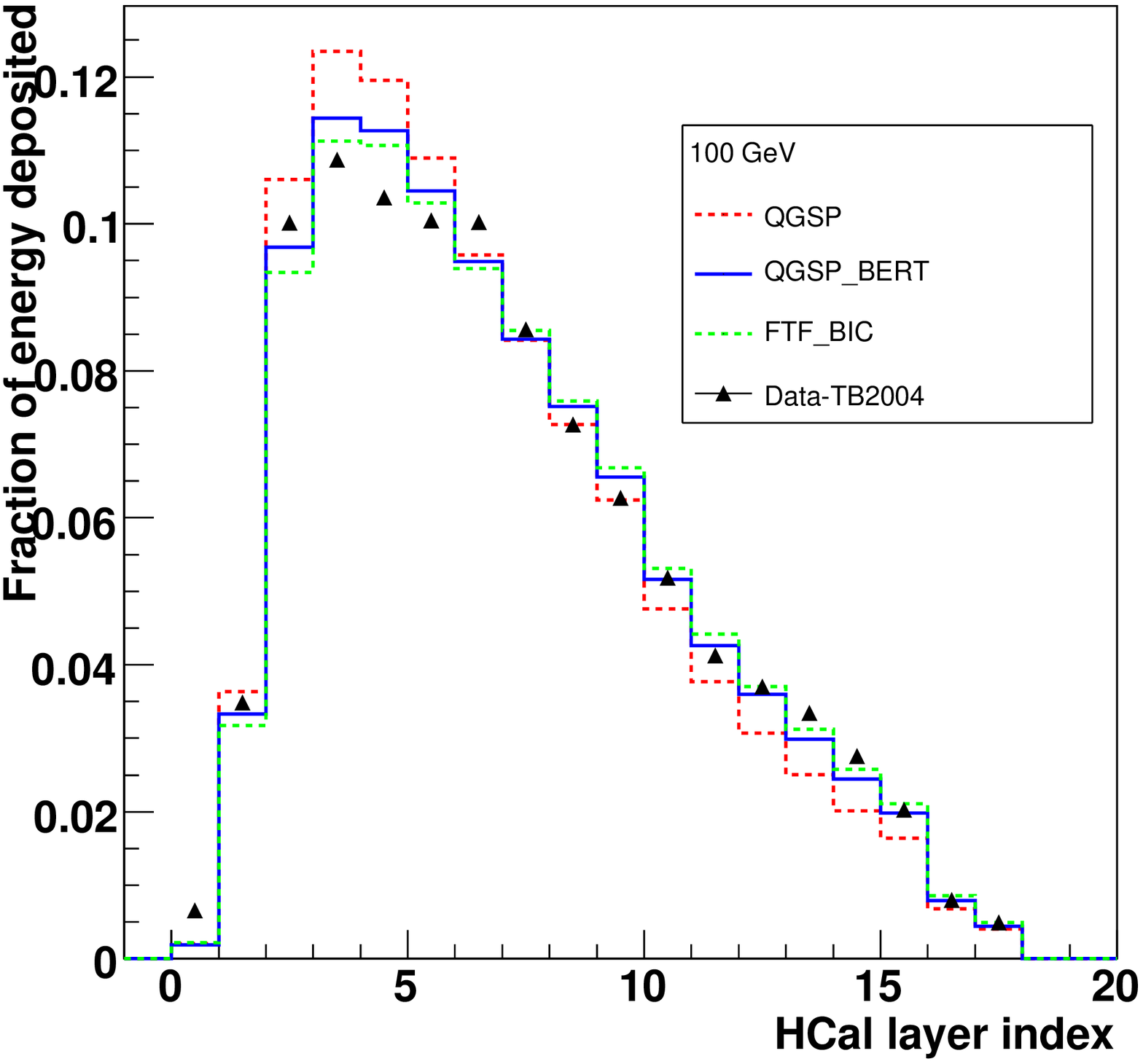}
&
  \includegraphics[width=0.21\linewidth]{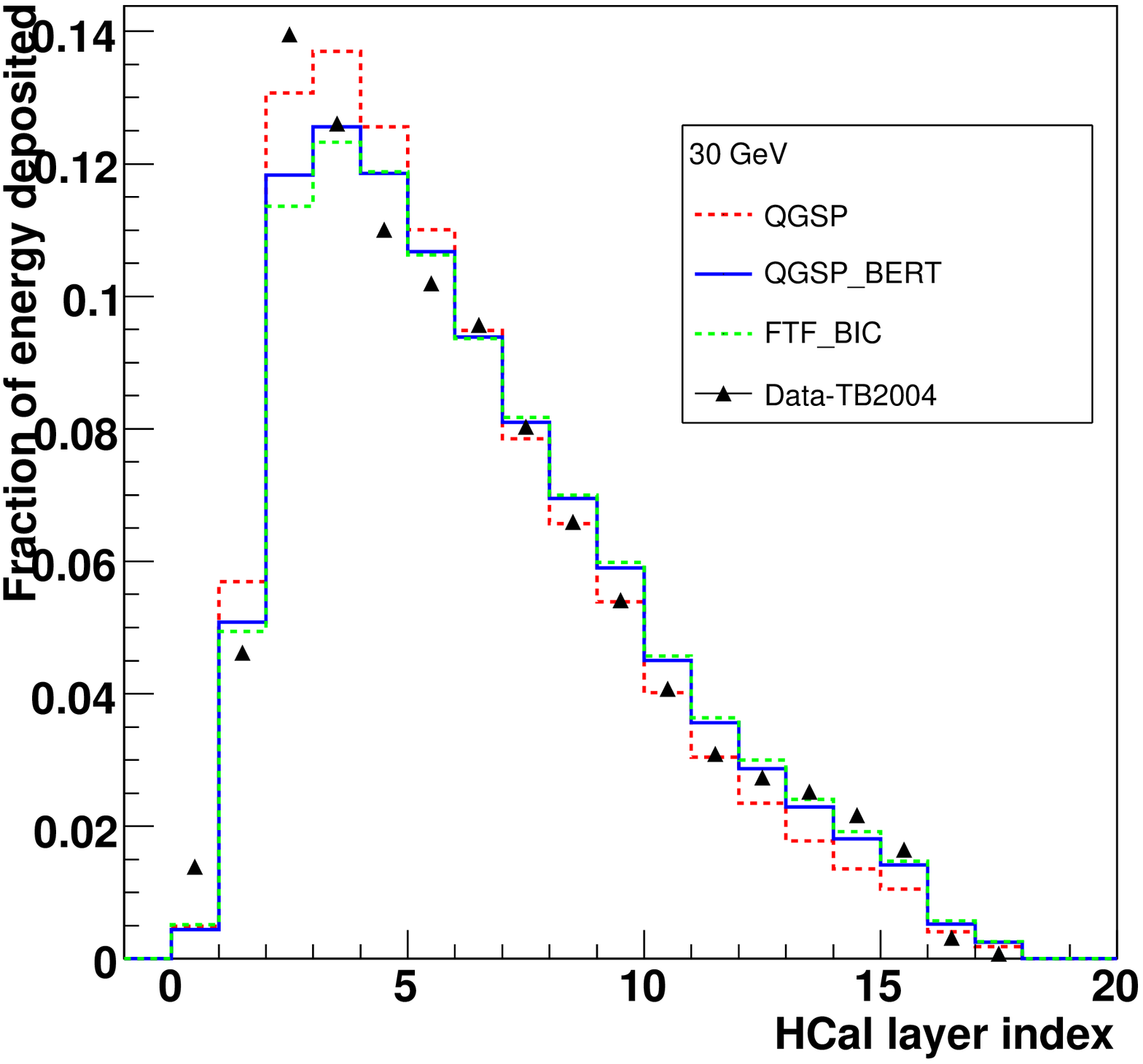}
 \\
a) & b) & c) \\
 \end{tabular}
 \caption{Longitudinal shower profiles for 300GeV (a); 100GeV (b); and 30GeV (c) pion beams compared to data.}
   \label{Long_shower}
\end{figure} 
\begin{figure}
 \begin{tabular}{ccc}
  \includegraphics[width=0.21\linewidth]{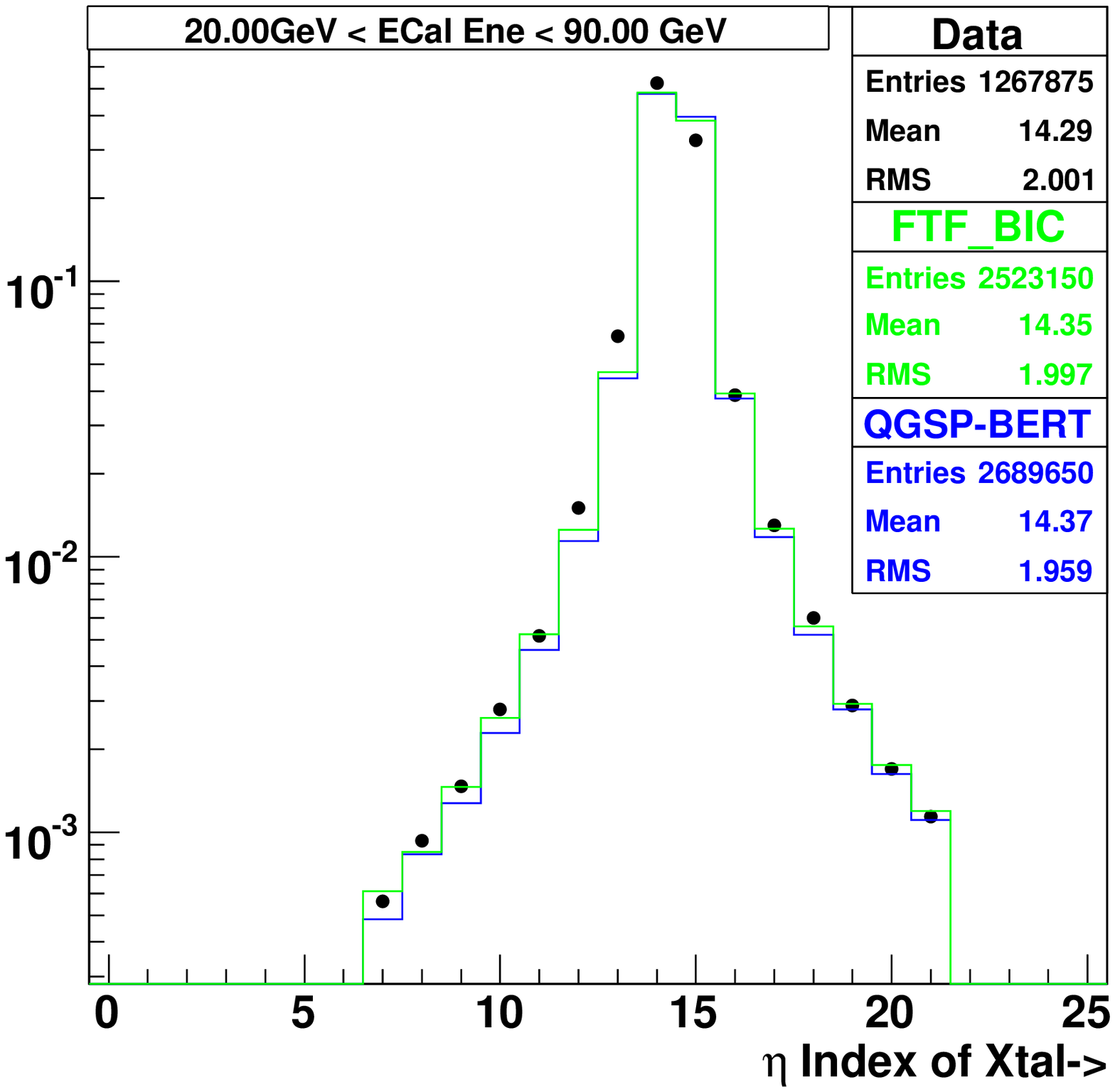}
&
  \includegraphics[width=0.21\linewidth]{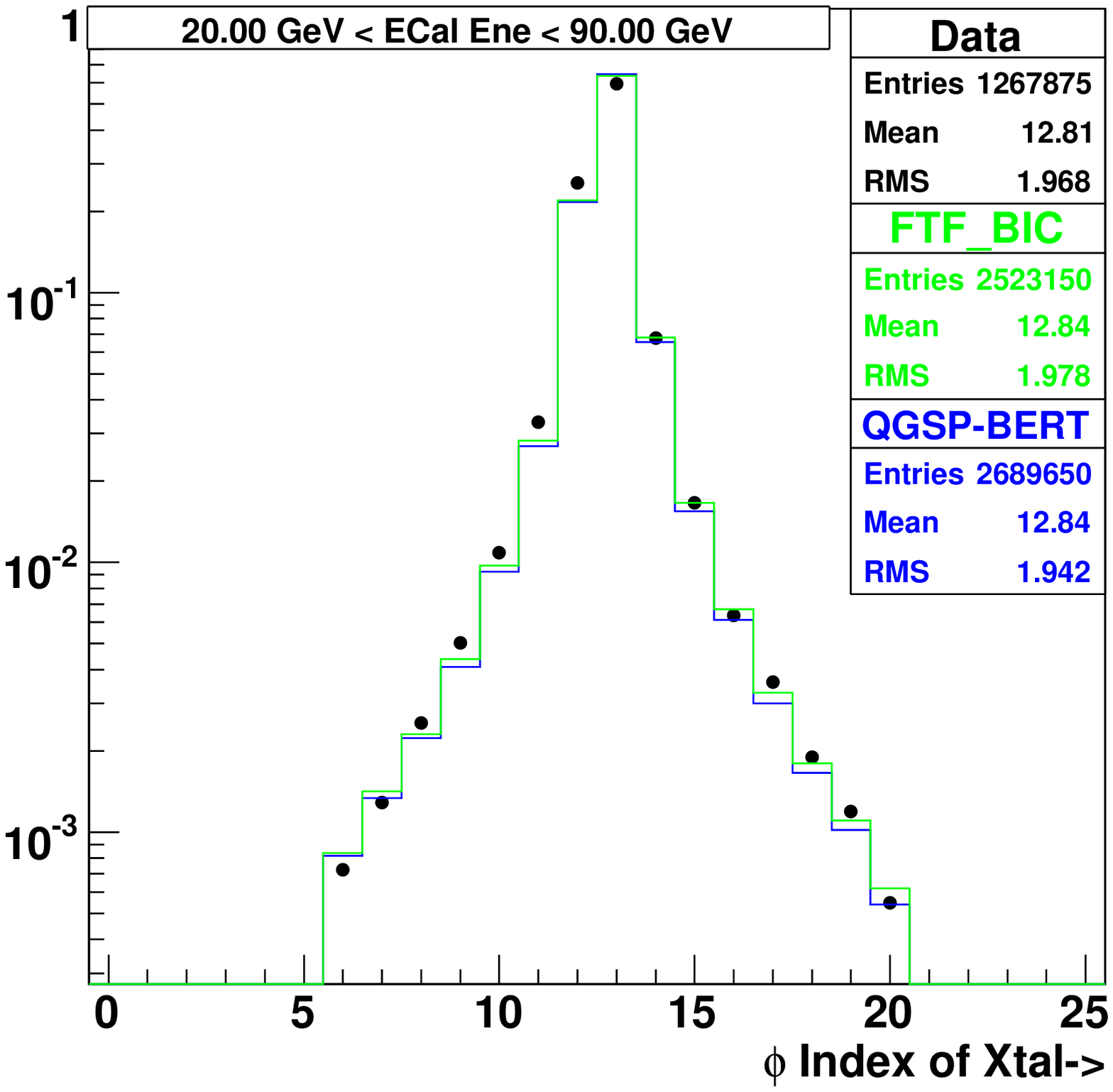}
&
  \includegraphics[width=0.21\linewidth]{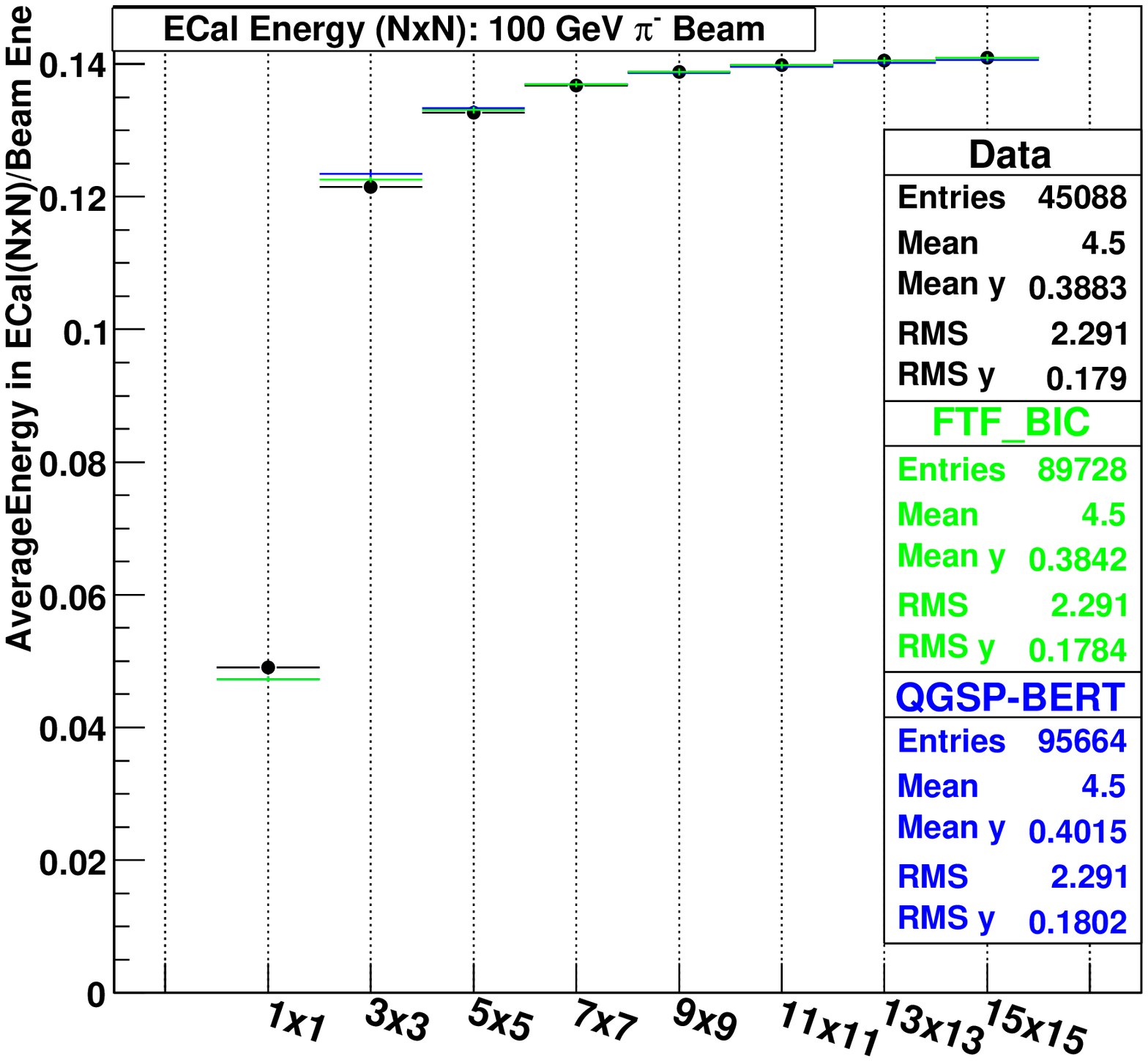}
 \\
a) & b) & c) \\
 \end{tabular}
 \caption{Comparison of ECAL Transverse shower profiles in $\eta$-direction (a) and $\phi$-direction (b) for 100GeV pions that interacted hadronically in the crystals; and shower containment of these showers (c) as a function of crystal matrix size (NxN), normalized to beam energy.}
   \label{ECAL_profile}
\end{figure}

\subsection{Shower Profiles}
Fig.\ref{Long_shower} shows the comparison of the longitudinal shower profiles measured in HB2 wedge and simulated with three different physics lists.
The QGSP-BERT physics list, using the Bertini-Cascade for low-energy interactions, shows best agreement with the measured profiles. 
Fig.\ref{ECAL_profile} compares the transverse shower profiles measured in ECAL, and the transverse shower containment. Agreement is good for both physics lists shown.
\begin{figure}
  \includegraphics[width=0.5\linewidth]{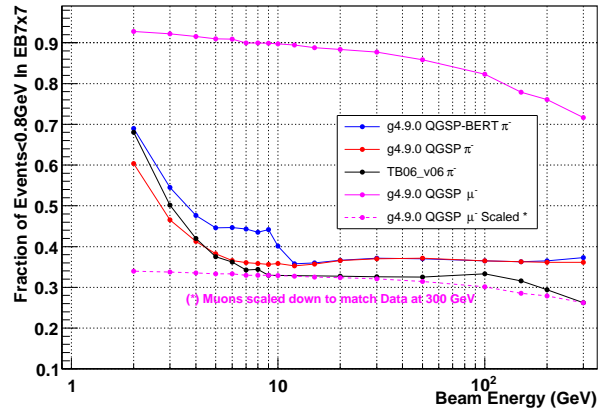}
 \caption{Fraction of pion events leaving MIP-like signal in ECAL, as a function of beam energy, compared to data. Superimposed is the same quantity for muon beam, and muon beam scaled down to match pion data at 300$GeV$}
   \label{MIP_fraction}
\end{figure}

\subsection{Fraction of MIP-like events}
An important measurement for the calibration of HCAL is the fraction of events which leave MIP-like signal in ECAL. As Fig.\ref{MIP_fraction} shows, the 
Bertini physics list has a major discrepancy at low energies, which is attributed to the inadequate description of the quasi-elastic processes. This is one of the major issues which the Geant4 team is addressing in the next releases of the software. 

\section{CONCLUSIONS}
Geant4 simulated response is in overall agreement with the measured data. We still observe small discrepancies in the following quantities:
\begin{itemize}
\item Simulated resolution is much better than the measured one, which is due to the smaller variation in signal distribution in HCAL at high energies.
\item Some discontinuities in the response curve at intermediate energies.
\item Fraction of MIP-like events in ECAL has peculiarities due to quasi-elastic processes in Bertini models.
\end{itemize}
We continue our active collaboration with the Geant4 team to fully understand and correct these discrepancies.

\end{document}